\documentclass[a4paper,11pt]{article}
\usepackage{pos}

\title{Multiwavelength Observations of the Blazar PKS~0735+178 in Spatial and Temporal Coincidence with an Astrophysical Neutrino Candidate IceCube-211208A}
 \ShortTitle{MWL Observations of the Blazar PKS~0735+178/IC211208A}

\manuallySeparateAuthors
\author*[a,b]{M. Pohl}
\author{ for the VERITAS collaboration,}
\author{ the H.E.S.S. collaboration, } 
\author[c]{and K.~Mori}

\affiliation[a]{Institute of Physics and Astronomy, University of Potsdam, 14476 Potsdam-Golm, Germany}

\affiliation[b]{DESY, Platanenallee 6, 15738 Zeuthen, Germany}

\affiliation[c]{Physics Department, Columbia University, New York, NY 10027, USA}

\emailAdd{marpohl@uni-potsdam.de}
\emailAdd{kaya@astro.columbia.edu}

\abstract{We report on multiwavelength target-of-opportunity observations of the blazar PKS 0735+178, located 2.2 degrees away from the best-fit position of the IceCube neutrino event 211208A. The source was in a high-flux state in the optical, ultraviolet, X-ray, and GeV gamma-ray bands around the time of the neutrino event, exhibiting daily variability in the soft X-ray flux. The X-ray data from Swift-XRT and NuSTAR characterize the transition between the low-energy and high-energy components of the broadband spectral energy distribution, and the gamma-ray data from Fermi-LAT, VERITAS, and H.E.S.S. require a spectral cut-off near 100 GeV. Both measurements provide strong constraints on leptonic and hadronic models. We analytically explore a synchrotron self-Compton model, an external Compton model, and a lepto-hadronic model. Models that are entirely based on internal photon fields face serious difficulties in matching the observed spectral energy distribution (SED). The existence of an external photon field in the source would instead explain the observed gamma-ray spectral cut-off in both leptonic and lepto-hadronic models, and it would allow a proton jet power that marginally agrees with the Eddington limit in the lepto-hadronic model. A numerical lepto-hadronic model with external target photons reproduces the observed SED and is reasonably consistent with the neutrino event despite requiring a high jet power.}

\ConferenceLogo{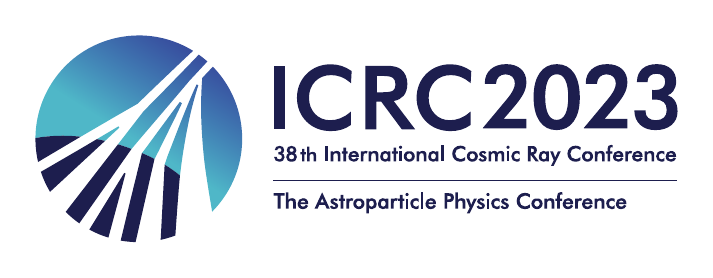}

\FullConference{%
38th International Cosmic Ray Conference (ICRC2023)\\
  26 July - 3 August, 2023\\
  Nagoya, Japan}

\begin{document}
\maketitle

\section{Introduction}
The distribution of arrival directions of the astrophysical neutrinos detected with the IceCube Neutrino Observatory \citep{IceCube13} suggests an extragalactic origin involving very many individual sources, besides a recently identified Galactic contribution \citep{doi:10.1126/science.adc9818}. Despite strong evidence for TeV neutrino emission from the nearby active galaxy {NGC~1068} \citep{Aartsen20,IceCube2022_NGC1068}, and the coincident detection in 2017 of a ~200-TeV neutrino event from TXS 0506+056 with temporally correlated gamma-ray flaring activity \citep{IceCube2018MMA}, there is no firm identification of the source class that produces the bulk of the diffuse neutrino emission in the TeV band.

Here we present results of multi-band observations of the blazar PKS~0735+178 contemporaneous with or immediately following the IceCube {astrophysical} neutrino candidate IceCube-211208A. An earlier study \cite{Sahakyan:2022nbz} explored the connection between PKS~0735+178 and IceCube-211208A on the basis of \textit{Fermi}-LAT, SWIFT (UVOT and XRT), and optical observations of {the blazar}. A similar study including NuSTAR data was performed recently \citep{Prince23}. We present, in addition, TeV-band data obtained with {the Very Energetic Radiation Imaging Telescope Array System (VERITAS)} and {the High Energy Stereoscopic System (H.E.S.S.)} that indicate a cut-off around 100 GeV in the gamma-ray spectrum. Acknowledging that the association between the neutrino event and the blazar PKS~0735+178 may well be spurious, we interpret the broadband SED of the source in the context of both leptonic and lepto-hadronic models, and discuss whether the neutrino event could have originated from the blazar. This proceedings paper provides a brief synopsis of the findings presented in an extended publication that will appear in the Astrophysical Journal \citep{acharyya2023multiwavelength}.

\section{Data}
IceCube-211208A was detected as a track-like event with an energy $E_\nu\approx 171$ TeV {and a 50.2\% probability of being astrophysical} \footnote{\url{https://gcn.gsfc.nasa.gov/notices_amon_g_b/136015_21306805.amon}} on December 8, 2021 \citep{2021GCN.31191....1I}. The {gamma-ray} blazar PKS 0735+178 (redshift $z=0.45$) {is located} immediately outside of the 90\% error region (2.13$^\circ$; statistical error only) for the neutrino event, 2.2$^\circ$ away from the best-fit position. Subsequent observations of PKS~0735+178 {revealed} flaring states in the radio band \citep{2021ATel15105....1K}, optical band \citep{2021ATel15098....1Z}, X-ray band \citep{2021ATel15102....1S,2021ATel15109....1D,2021ATel15113....1F}, and GeV gamma-ray band \citep{2021ATel15099....1G}. 

We triggered NuSTAR {observations} on December 11 and 13, 2021, and found the X-ray spectrum to be harder than that seen with SWIFT-XRT \citep{2021ATel15113....1F}. In addition, VERITAS and H.E.S.S. performed ToO observations that yielded upper limits above {100~GeV}. We present the SED in Figure~\ref{fig1}. The VERITAS upper limit at $330$~GeV is {particularly relevant. After correction for absorption by extragalactic background light (EBL) \citep{Dominguez11}  for the nominal redshift $z=0.45$ \citep{2012A&A...547A...1N}, the flux limit is} about a factor of ten {below the log-parabola extrapolation of the \textit{Fermi}-LAT spectrum}. This finding suggests that either the redshift is misestimated, or there is an intrinsic cut-off at 100~GeV in the emission spectrum, or the source becomes optically thick at 100~GeV, on account of pair production on ambient hard-X-ray photons.  

The synchrotron peak frequency is loosely constrained by the rather flat optical/UV spectrum. Nominally, the highest energy flux is observed at a few $10^{14}$~Hz, or abourt 1~eV, with UVOT data suggesting an optical/near-UV flux that is a slowly declining function of frequency. Beyond the near-UV band, there are no measurements up to $300$~eV, where XRT data indicate a flux a factor of 30 below that at $3$~eV, indicating that the cut-off energy is below $100$~eV, and likely far lower based on the slowly declining UV spectrum.  
The measured soft and hard X-ray spectra fully constrain the tail of the synchrotron emission and the beginning of the high-energy component of the SED, respectively. The transition occurs at a few keV. 

Day-scale variability in the GeV-band gamma-ray flux is evident, but there is no evidence of spectral variability. The soft X-ray flux exhibited daily variability, but no hard X-ray variability was observed between the two NuSTAR observations on December 11 and 13. Given the daily variability, the SED on December 13 is the focus of our modeling.

\section{Discussion}

The radius of the emission region is constrained by the soft X-ray variability, and to a lesser degree by that in gamma-rays, to $R\lesssim 10^{16}\ \mathrm{cm}$. The size estimate permits turning the observed synchrotron flux at the $\nu F_\nu$ peak into the photon energy density in the emission zone, 
\begin{equation}
U'=\epsilon' {n_{\ln\epsilon}'}\approx
\frac{5\times 10^{-3}\ \mathrm{erg\,cm^{-3}}}
{D_{25}^4\,R_{16}^2}\, ,
\label{eq:u_eps}
\end{equation}
where we denote the radius of the emission zone as $R' = R_{16}\,(10^{16}\ \mathrm{cm})$ and the Doppler factor as $D=25\,D_{25}$. Primed quantities are measured in the jet frame, and $n_{\ln\epsilon}$ stands for the photon density per logarithmic energy interval. This is the photon field that Inverse-Comption emission and potentially neutrino emission is produced with.

\begin{figure}
\includegraphics[width=.95\textwidth]{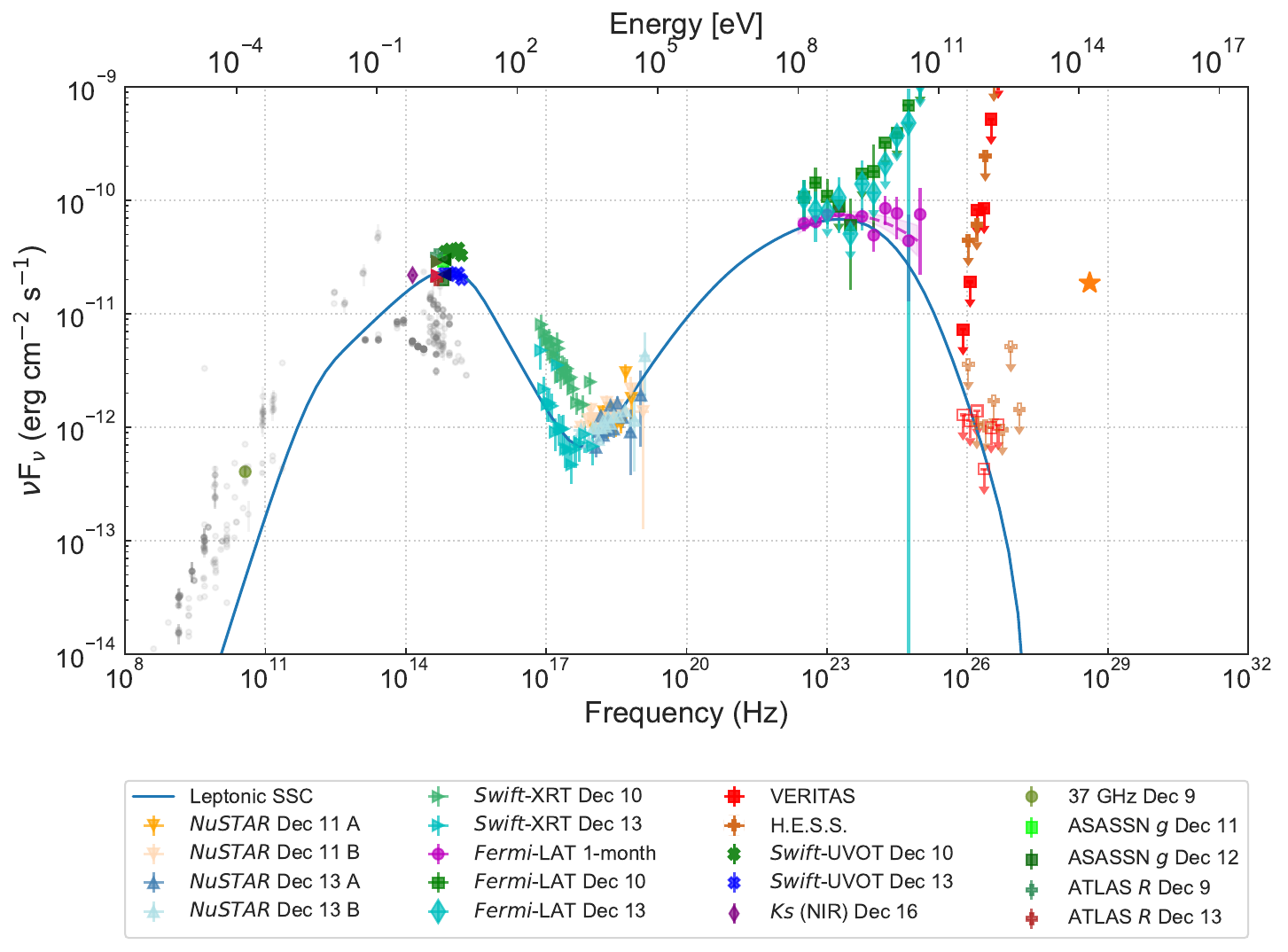}
\caption{The broadband SED and a best-fit one-zone SSC model. For the TeV-band data, open symbols indicate the measured flux and filled symbols refer to the de-absorbed flux. Flux points in grey reflect archival data. The star at 170~TeV marks the neutrino flux corresponding to one observed event in eight months of observing time.}
\label{fig1}
\end{figure}

The synchrotron spectrum {seems to roll over} at a few $10^{14}$~Hz in frequency, and we write the peak photon energy as $\epsilon_\mathrm{peak}=\epsilon_\mathrm{peak,eV}$~eV. There {has} {to} be {a Synchrotron-self-Compton (SSC) contribution} to the SED. Since the Thomson limit applies, the ratio between the inverse-Compton (IC) and the synchrotron peak frequencies ($100\ \mathrm{GeV}/\epsilon_\mathrm{peak}$) gives the square of the peak Lorentz factor of electrons, leading to 
\begin{equation}
\gamma_\mathrm{peak}\approx \frac{3\times 10^5}{\sqrt{\epsilon_\mathrm{peak,eV}}}\ .
\label{eq:1}
\end{equation}
The synchrotron peak frequency and the peak Lorentz factor of electrons constrain the magnetic-field strength in the emission zone,
\begin{equation}
B' D\approx ( 10^{-3}\ \mathrm{G})\,\epsilon_\mathrm{peak,eV}^2\ .
\label{eq:BD}
\end{equation}
The optical/UV synchrotron emission appears to be roughly as bright in $\nu F_\nu$ as is the GeV-scale gamma-ray emission, indicating that the energy densities in the photon field and in the magnetic field must be comparable. That requires a very high cut-off frequency of the synchrotron emission or an emission zone that is very much larger than $10^{16}\ \mathrm{cm}$. As the UVOT data suggest $\epsilon_\mathrm{peak,eV}\approx 1$, and the source radius cannot be made arbitrarily large, on account of the day-scale variability, the SSC model has difficulty reproducing a 100-GeV cut-off in the gamma-ray spectrum. We present a best-fit SSC model and the observed SED in Fig.~\ref{fig1}. To be noted from the figure is the deficit in the SSC spectrum of gamma rays around 10~GeV.

Considering an internal photon field with energy density as given in Eq.~\ref{eq:u_eps}, p-$\gamma$ interactions of PeV-scale protons can yield a neutrino flux commensurate with one detected neutrino per month only, if the proton acceleration power, and hence the jet luminosity, vastly exceed the Eddington luminosity  \citep{2002AJ....123.2352X}. 

The featureless optical spectrum of PKS~0735+178 suggests that it may be a BL Lac. It has been noted in BL Lacs that the jet may pass through a region harboring a significant jet-external photon field \citep{2022ApJ...926...95F}. In the jet frame the external photons would be in the keV band and have an energy density much higher than that of the synchrotron photons. If only 1\% 
of the observed soft-UV flux were rescattered in a pc-scale central region, then one neutrino event in a month of high source activity would be in marginal agreement with the Eddington limit. We performed a numerical analysis using a state-of-the-art code \citep{Cerruti15} and confirmed the analytically derived findings that are presented above.

\section{Summary}
Based on extensive follow-up observations across the electromagnetic spectrum, we analysed the state of the blazar PKS~0735+178 in the context of a potential association with a VHE neutrino event that was seen with IceCube in December 2021. We analytically demonstrated that the observed SED of the blazar, {especially the gamma-ray cutoff in the TeV data,} constitutes a challenge to a simple one-zone SSC model. It may be explained by an SSC/EC scenario, which naturally provides the observed 100-GeV cut-off through Klein-Nishina suppression and $\gamma$-$\gamma$ pair absorption. The potential association with the neutrino event could be explained with a lepto-hadronic scenario that likewise must involve an external photon field, otherwise there would be no agreement between the Eddington limit on the {jet} luminosity and the observed IceCube neutrino rate. 

%
\acknowledgments
This research is supported by grants from the U.S. Department of Energy Office of Science, the U.S. National Science Foundation and the Smithsonian Institution, by NSERC in Canada, and by the Helmholtz Association in Germany. This research used resources provided by the Open Science Grid, which is supported by the National Science Foundation and the U.S. Department of Energy's Office of Science, and resources of the National Energy Research Scientific Computing Center (NERSC), a U.S. Department of Energy Office of Science User Facility operated under Contract No. DE-AC02-05CH11231. We acknowledge the excellent work of the technical support staff at the Fred Lawrence Whipple Observatory and at the collaborating institutions in the construction and operation of the instrument. 

This work was supported by NASA grants 80NSSC22K0573, {80NSSC22K1515, 80NSSC22K0950, 80NSSC20K1587, 80NSSC20K1494,} and NSF grant PHY-1806554. 

The support of the Namibian authorities and of the University of
Namibia in facilitating the construction and operation of H.E.S.S.
is gratefully acknowledged, as is the support by the German
Ministry for Education and Research (BMBF), the Max Planck Society,
the German Research Foundation (DFG), the Helmholtz Association,
the Alexander von Humboldt Foundation, the French Ministry of
Higher Education, Research and Innovation, the Centre National de
la Recherche Scientifique (CNRS/IN2P3 and CNRS/INSU), the
Commissariat à l’énergie atomique et aux énergies alternatives
(CEA), the U.K. Science and Technology Facilities Council (STFC),
the Irish Research Council (IRC) and the Science Foundation Ireland
(SFI), the Knut and Alice Wallenberg Foundation, the Polish
Ministry of Education and Science, agreement no. 2021/WK/06, the
South African Department of Science and Technology and National
Research Foundation, the University of Namibia, the National
Commission on Research, Science \& Technology of Namibia (NCRST),
the Austrian Federal Ministry of Education, Science and Research
and the Austrian Science Fund (FWF), the Australian Research
Council (ARC), the Japan Society for the Promotion of Science, the
University of Amsterdam and the Science Committee of Armenia grant
21AG-1C085. We appreciate the excellent work of the technical
support staff in Berlin, Zeuthen, Heidelberg, Palaiseau, Paris,
Saclay, Tübingen and in Namibia in the construction and operation
of the equipment. This work benefited from services provided by the
H.E.S.S. Virtual Organisation, supported by the national resource
providers of the EGI Federation.

This work made use of data supplied by the UK SWIFT Science Data Centre at the University of Leicester.

This work has made use of data from the Asteroid Terrestrial-impact Last Alert
System (ATLAS) project. The Asteroid Terrestrial-impact Last Alert System
(ATLAS) project is primarily funded to search for near earth asteroids through
NASA grants NN12AR55G, 80NSSC18K0284, and 80NSSC18K1575; byproducts of
the NEO search include images and catalogs from the survey area. This work was
partially funded by Kepler/K2 grant J1944/80NSSC19K0112 and HST GO-15889,
and STFC grants ST/T000198/1 and ST/S006109/1. The ATLAS science products
have been made possible through the contributions of the University of Hawaii
Institute for Astronomy, the Queen’s University Belfast, the Space Telescope
Science Institute, the South African Astronomical Observatory, and The
Millennium Institute of Astrophysics (MAS), Chile.

\bibliographystyle{JHEP}
\bibliography{0735+178}

\clearpage

\section*{Full Author List: VERITAS Collaboration}

\scriptsize
\noindent
A.~Acharyya$^{1}$,
C.~B.~Adams$^{2}$,
A.~Archer$^{3}$,
P.~Bangale$^{4}$,
J.~T.~Bartkoske$^{5}$,
P.~Batista$^{6}$,
W.~Benbow$^{7}$,
J.~L.~Christiansen$^{8}$,
A.~J.~Chromey$^{7}$,
A.~Duerr$^{5}$,
M.~Errando$^{9}$,
Q.~Feng$^{7}$,
G.~M.~Foote$^{4}$,
L.~Fortson$^{10}$,
A.~Furniss$^{11, 12}$,
W.~Hanlon$^{7}$,
O.~Hervet$^{12}$,
C.~E.~Hinrichs$^{7,13}$,
J.~Hoang$^{12}$,
J.~Holder$^{4}$,
Z.~Hughes$^{9}$,
T.~B.~Humensky$^{14,15}$,
W.~Jin$^{1}$,
M.~N.~Johnson$^{12}$,
M.~Kertzman$^{3}$,
M.~Kherlakian$^{6}$,
D.~Kieda$^{5}$,
T.~K.~Kleiner$^{6}$,
N.~Korzoun$^{4}$,
S.~Kumar$^{14}$,
M.~J.~Lang$^{16}$,
M.~Lundy$^{17}$,
G.~Maier$^{6}$,
C.~E~McGrath$^{18}$,
M.~J.~Millard$^{19}$,
C.~L.~Mooney$^{4}$,
P.~Moriarty$^{16}$,
R.~Mukherjee$^{20}$,
S.~O'Brien$^{17,21}$,
R.~A.~Ong$^{22}$,
N.~Park$^{23}$,
C.~Poggemann$^{8}$,
M.~Pohl$^{24,6}$,
E.~Pueschel$^{6}$,
J.~Quinn$^{18}$,
P.~L.~Rabinowitz$^{9}$,
K.~Ragan$^{17}$,
P.~T.~Reynolds$^{25}$,
D.~Ribeiro$^{10}$,
E.~Roache$^{7}$,
J.~L.~Ryan$^{22}$,
I.~Sadeh$^{6}$,
L.~Saha$^{7}$,
M.~Santander$^{1}$,
G.~H.~Sembroski$^{26}$,
R.~Shang$^{20}$,
M.~Splettstoesser$^{12}$,
A.~K.~Talluri$^{10}$,
J.~V.~Tucci$^{27}$,
V.~V.~Vassiliev$^{22}$,
A.~Weinstein$^{28}$,
D.~A.~Williams$^{12}$,
S.~L.~Wong$^{17}$,
and
J.~Woo$^{29}$\\
\\
\noindent
$^{1}${Department of Physics and Astronomy, University of Alabama, Tuscaloosa, AL 35487, USA}

\noindent
$^{2}${Physics Department, Columbia University, New York, NY 10027, USA}

\noindent
$^{3}${Department of Physics and Astronomy, DePauw University, Greencastle, IN 46135-0037, USA}

\noindent
$^{4}${Department of Physics and Astronomy and the Bartol Research Institute, University of Delaware, Newark, DE 19716, USA}

\noindent
$^{5}${Department of Physics and Astronomy, University of Utah, Salt Lake City, UT 84112, USA}

\noindent
$^{6}${DESY, Platanenallee 6, 15738 Zeuthen, Germany}

\noindent
$^{7}${Center for Astrophysics $|$ Harvard \& Smithsonian, Cambridge, MA 02138, USA}

\noindent
$^{8}${Physics Department, California Polytechnic State University, San Luis Obispo, CA 94307, USA}

\noindent
$^{9}${Department of Physics, Washington University, St. Louis, MO 63130, USA}

\noindent
$^{10}${School of Physics and Astronomy, University of Minnesota, Minneapolis, MN 55455, USA}

\noindent
$^{11}${Department of Physics, California State University - East Bay, Hayward, CA 94542, USA}

\noindent
$^{12}${Santa Cruz Institute for Particle Physics and Department of Physics, University of California, Santa Cruz, CA 95064, USA}

\noindent
$^{13}${Department of Physics and Astronomy, Dartmouth College, 6127 Wilder Laboratory, Hanover, NH 03755 USA}

\noindent
$^{14}${Department of Physics, University of Maryland, College Park, MD, USA }

\noindent
$^{15}${NASA GSFC, Greenbelt, MD 20771, USA}

\noindent
$^{16}${School of Natural Sciences, University of Galway, University Road, Galway, H91 TK33, Ireland}

\noindent
$^{17}${Physics Department, McGill University, Montreal, QC H3A 2T8, Canada}

\noindent
$^{18}${School of Physics, University College Dublin, Belfield, Dublin 4, Ireland}

\noindent
$^{19}${Department of Physics and Astronomy, University of Iowa, Van Allen Hall, Iowa City, IA 52242, USA}

\noindent
$^{20}${Department of Physics and Astronomy, Barnard College, Columbia University, NY 10027, USA}

\noindent
$^{21}${ Arthur B. McDonald Canadian Astroparticle Physics Research Institute, 64 Bader Lane, Queen's University, Kingston, ON Canada, K7L 3N6}

\noindent
$^{22}${Department of Physics and Astronomy, University of California, Los Angeles, CA 90095, USA}

\noindent
$^{23}${Department of Physics, Engineering Physics and Astronomy, Queen's University, Kingston, ON K7L 3N6, Canada}

\noindent
$^{24}${Institute of Physics and Astronomy, University of Potsdam, 14476 Potsdam-Golm, Germany}

\noindent
$^{25}${Department of Physical Sciences, Munster Technological University, Bishopstown, Cork, T12 P928, Ireland}

\noindent
$^{26}${Department of Physics and Astronomy, Purdue University, West Lafayette, IN 47907, USA}

\noindent
$^{27}${Department of Physics, Indiana University-Purdue University Indianapolis, Indianapolis, IN 46202, USA}

\noindent
$^{28}${Department of Physics and Astronomy, Iowa State University, Ames, IA 50011, USA}

\noindent
$^{29}${Columbia Astrophysics Laboratory, Columbia University, New York, NY 10027, USA}

\section*{Full Author List: H.E.S.S. Collaboration}

\scriptsize
\noindent
F.~Aharonian$^{1,2,3}$, 
F.~Ait~Benkhali$^{4}$, 
A.~Alkan$^{5}$, 
J.~Aschersleben$^{6}$, 
H.~Ashkar$^{7}$, 
M.~Backes$^{8,9}$, 
A.~Baktash$^{10}$, 
V.~Barbosa~Martins$^{11}$, 
A.~Barnacka$^{12}$, 
J.~Barnard$^{13}$
% ...

\medskip

\noindent
$^{1}$Dublin Institute for Advanced Studies, 31 Fitzwilliam Place, Dublin 2, Ireland\\
$^{2}$Max-Planck-Institut f\"ur Kernphysik, P.O. Box 103980, D 69029 Heidelberg, Germany\\
$^{3}$Yerevan State University,  1 Alek Manukyan St, Yerevan 0025, Armenia\\
$^{4}$Landessternwarte, Universit\"at Heidelberg, K\"onigstuhl, D 69117 Heidelberg, Germany\\
$^{5}$IRFU, CEA, Universit\'e Paris-Saclay, F-91191 Gif-sur-Yvette, France\\
$^{6}$Kapteyn Astronomical Institute, University of Groningen, Landleven 12, 9747 AD Groningen, The Netherlands\\
$^{7}$Laboratoire Leprince-Ringuet, École Polytechnique, CNRS, Institut Polytechnique de Paris, F-91128 Palaiseau, France\\
$^{8}$University of Namibia, Department of Physics, Private Bag 13301, Windhoek 10005, Namibia\\
$^{9}$Centre for Space Research, North-West University, Potchefstroom 2520, South Africa\\
$^{10}$Universit\"at Hamburg, Institut f\"ur Experimentalphysik, Luruper Chaussee 149, D 22761 Hamburg, Germany\\
$^{11}$DESY, D-15738 Zeuthen, Germany\\
$^{12}$Obserwatorium Astronomiczne, Uniwersytet Jagiello{\'n}ski, ul. Orla 171, 30-244 Krak{\'o}w, Poland\\
$^{13}$Department of Physics, University of the Free State,  PO Box 339, Bloemfontein 9300, South Africa\\
% ...

\end{document}